\documentclass[
superscriptaddress,
 amsmath,amssymb,
 aps,
prl,
 twocolumn
]{revtex4-2}

\usepackage{graphicx}
\usepackage{dcolumn}
\usepackage{bm}
\usepackage{color}

\usepackage{physics}

\newcommand{\un}{\bm{\widehat{\sigma}}}
\newcommand{\e}{ \mathbf{e}} 
\newcommand{\vv}{ \mathbf{v}} 
\newcommand{\bop}{\widehat{b}_{\boldsymbol{\sigma}}}

\begin{document}


\title{
Cooling mechanism controls motility-induced phase separation in inertial active liquids 
}

\author{Manuel Mayo}
\affiliation{F\'\i sica Te\'orica, Universidad de Sevilla, Apartado de Correos 1065, E-41080 Sevilla, Spain}

\author{Lorenzo Caprini}
\affiliation{Sapienza University of Rome, Physics Department, P.le A. Moro 2, Rome, Italy}

\author{Mar\'\i a Isabel Garc\'\i a de Soria} 
\affiliation{F\'\i sica Te\'orica, Universidad de Sevilla, Apartado de Correos 1065, E-41080 Sevilla, Spain}

\author{Umberto Marini Bettolo Marconi}
\affiliation{School of Sciences and Technology, University of Camerino, Via Madonna delle Carceri, Italy}

\author{Pablo Maynar}
\affiliation{F\'\i sica Te\'orica, Universidad de Sevilla, Apartado de Correos 1065, E-41080 Sevilla, Spain}

\author{Luca Pizzoli}
\affiliation{Sapienza University of Rome, Physics Department, P.le A. Moro 2, Rome, Italy}

\author{Andrea Puglisi}
\email{andrea.puglisi@cnr.it}
\affiliation{Istituto dei Sistemi Complessi - Consiglio Nazionale delle Ricerche, P.le A. Moro 2, Rome Italy}
\affiliation{Sapienza University of Rome, Physics Department, P.le A. Moro 2, Rome, Italy}
\affiliation{INFN, Sezione Roma2, Via della Ricerca Scientifica 1, I-00133, Rome, Italy}    

\date{\today}

\begin{abstract}
Motility-induced phase separation (MIPS) is a central collective phenomenon in active matter, theoretically established in the overdamped regime. We discover that the dynamical origin of MIPS is fundamentally altered by inertia, which induces a cooling mechanism absent in overdamped active matter. This conclusion is supported by an active variant of the direct simulation Monte Carlo method and by a kinetic theory for inertial self-propelled hard spheres derived from the microscopic dynamics. In contrast to the overdamped case, both analyses demonstrate that inertial MIPS does not rely on volume exclusion but on a cooling mechanism involving density, polarization, and temperature fields. This mechanism emerges from the competition between activity and a density dependent collision rate, arising from spatial correlations between colliding particles. These findings open a pathway to fundamentally connect inertial active matter with granular physics.

\end{abstract}

\maketitle

{\paragraph{Introduction --} 

Active matter~\cite{marchetti2013hydrodynamics,Elgeti2015,bechinger2016active} encompasses a broad class of systems: At submicron scales, it includes intracellular proteins, filaments, and organelles~\cite{winkler2020physics}. On the micron scale, active systems range from sperm cells~\cite{nath2023microfluidic} and bacteria~\cite{di2010bacterial,arlt2018painting} to synthetic active colloids~\cite{buttinoni2013dynamical,bricard2013emergence}.
Active matter is also ubiquitous at macroscopic scales and can be encountered in everyday life, from animal groups~\cite{cavagna2014bird} and pedestrian crowds~\cite{moussaid2010walking} to granular robots~\cite{leyman2018tuning,agrawal2020scale} and drone swarms~\cite{vasarhelyi2018optimized}. The defining feature of these systems is an internal mechanism of energy injection that drives them far from thermodynamic equilibrium~\cite{o2022time}, for example by sustaining self-propelled motion.


Active materials can self-organize even under conditions where passive matter remains disordered. A starking example is motility-induced phase separation (MIPS)~\cite{cates2015motility,gonnella2015motility,bialke2015active}, a nonequilibrium coexistence between dense and dilute phases generated by the competition between self-propulsion and volume exclusion~\cite{fily2012athermal,Redner2013Structure,levis2017active,digregorio2018full,hermann2019non}. These two mechanisms cause active particles moving at constant speed to block each other inducing cluster nucleation without attractions~\cite{klamser2018thermodynamic,caprini2020spontaneous,broker2023orientation,garcia2025dynamics}.  MIPS has been extensively analyzed using nonequilibrium statistical mechanics, particularly in the overdamped regime, where kinetic energy rapidly relaxes to the bath temperature~\cite{speck2016collective,wittmann2017effective}. In this limit, MIPS can be described through an effective free-energy approach with a modified Maxwell construction~\cite{solon2015pressure}, or via mechanical~\cite{omar2023mechanical} and kinetic theories~\cite{soto2024kinetic}.


Understanding the role of inertia has become increasingly important, given the large variety of active systems observed at macroscopic scales~\cite{lowen2020inertial}, including polar granular particles powered by internal motors~\cite{baconnier2022selective,siebers2023exploiting,engbring2023nonlinear,casiulis2024geometric}, grains driven by vibrating plates~\cite{aranson2007swirling,kudrolli2008,deseigne2010collective,kumar2014flocking,Koumakis2016,antonov2024inertial}, granular spinners~\cite{scholz2018rotating,workamp2018symmetry,lopez2021pseudo,lopez2022chirality}, whirling fruits~\cite{rabault2019curving}, and flying beetles~\cite{mukundarajan2016surface}, typically operating in air where the Stokes time is on the order of seconds. In these systems, inertia cannot be neglected and influences both single-particle dynamics~\cite{takatori2017inertial,scholz2018inertial,leoni2020surfing,caprini2021inertial,nguyen2021active} and collective behavior~\cite{de2022collective,kryuchkov2023inertia,deblais2018boundaries,antonov2025self}.
%
In particular, inertia suppresses MIPS~\cite{caprini2024dynamical,mandal2019motility,de2022motility} and, through a bounce-back mechanism~\cite{horvath2023bouncing,caprini2024emergent}, generates a temperature difference between the dense cluster and the dilute phase~\cite{mandal2019motility} -- a feature recently exploited to propose a nonequilibrium refrigerator~\cite{hecht2022active}. Numerical studies also reveal oscillatory behaviors~\cite{dai2020phase} and a nucleation-like MIPS~\cite{su2021inertia}, in contrast to the spinodal decomposition typical of overdamped active systems~\cite{cates2015motility}. These findings indicate that the mechanism underlying MIPS may differ fundamentally when inertia dominates as an effect of the temperature field. However, beyond a mechanical theory predicting the coexistence line~\cite{feng2025theory}, a comprehensive kinetic theory is still lacking, leaving open questions about the role of the temperature.

Here, we demonstrate that MIPS in inertial active systems originates from a cooling mechanism absent in the overdamped regime. 
This mechanism can be understood monitoring a binary collision between two particles (Fig.~\ref{fig:sketch}(d)): after the impact, the particle velocity is reflected while its self-propulsion remains unchanged. As a consequence, the efficiency of self-propulsion (see Eq.~\eqref{eq:kinen} below) is typically reduced, effectively lowering the kinetic energy and leading to a pressure inversion reminiscent of clustering in granular materials~\cite{goldhirsch1993clustering,maynar2019understanding}. The cooling origin of inertial MIPS is confirmed by simulations performed with an active version of the Direct Monte Carlo scheme (Bird algorithm) where the volume exclusion can be switched off. These findings are supported by a kinetic theory derived from the Boltzmann–Fokker–Planck equation for the microscopic dynamics, which predicts the onset of a linear instability driven by activity, inertia, and the Enskog correction to Molecular Chaos.

\begin{figure}
    {\centering
        \includegraphics[width=0.99\linewidth]{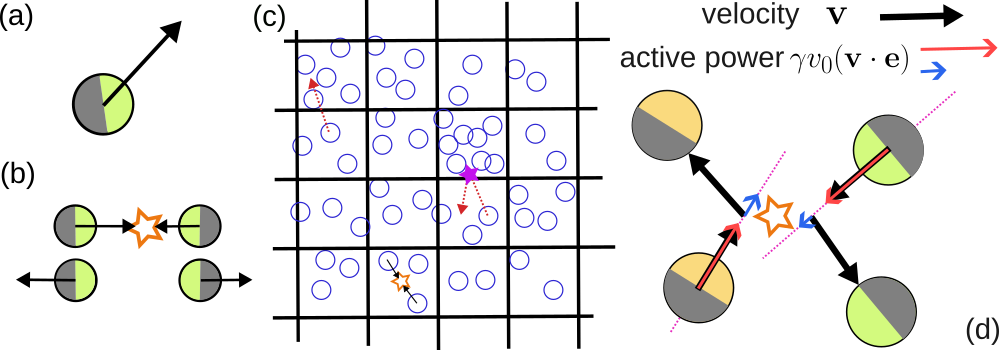}}
    \caption{Inertia-induced cooling in active systems. (a) Particles are endowed with both a velocity ${\bf v}$ (black arrow) and an active force $\gamma v_0 \mathbf{e}$ directed along the direction normal to the hemisphere, going from gray to color. (b) After a collision, velocities are instantaneously changed while active forces remains unchanged. (c) In ADSMC, time and space are discretized. At each time step the first phase is free motion (red arrow), the second phase is interaction: pair collisions in the same cell (orange star), and (when implemented) ``excluded volume events'' of single particles with overcrowded cells (purple star). (d) Collisional cooling effect. Cap colors identify the same particles before and after the collision. The modulus of the colored arrows represents the "active power" $\gamma v_{0} (\mathbf{v} \cdot \mathbf{e})$ (in $d=2$), which decreases after impact (from red to blue). Since $\gamma v_{0} (\mathbf{v} \cdot \mathbf{e})$ determines the value of the local temperature field, see Eq.~\eqref{eq:kinen}, inertial active particles undergo an effective cooling mechanism during collisions.  
    }
    \label{fig:sketch}
\end{figure}
\vspace{.3cm} 

{\paragraph{Model} --} We consider $N$ active spheres of  mass $m$ and diameter $\sigma$,  with positions ${\bf r}_i$, and velocities  $\dot{\bf r}_i={\bf v}_i$  evolving as
\begin{equation}
\label{eq:dynamics_v}
m\dot{\bf v}_i={\bf f}_i^{int}+\gamma v_0{\bf e}_i-\gamma {\bf v}_i+\sqrt{2 \gamma T_0}{\pmb \eta}_i(t)\,,
\end{equation}
where ${\bf f}^{int}_i$ denotes interparticle interactions and $\gamma v_0{\bf e_i}$ is the self-propulsion force (Fig.~\ref{fig:sketch}(a)). The parameters $\gamma$ and $T_0$ represent the drag coefficient and bath temperature, respectively, with Boltzmann’s constant set to $k_B=1$. Here ${\pmb \eta}_i(t)$ is a white noise vector with zero mean and unit variance.
The self-propulsion vector ${\bf e}_i$ follows an active Ornstein–Uhlenbeck particle (AOUP) dynamics with autocorrelation time $\tau$:
\begin{equation}
\label{eq:dynamics_f}
\tau\dot{\mathbf{e}}_i = -\mathbf{e}_i+ \sqrt{2\tau}\boldsymbol{\xi}_i \,,
\end{equation}
where $\boldsymbol{\xi}_i$ is a white noise vector with zero mean and unit variance, implying $\langle \mathbf{e}_i^2\rangle=d$ where $d$ is the space dimensionality. 
Two relevant dimensionless parameters are the Péclet number $\ell=\tau v_0/\sigma$ and the Stokes number $\textrm{St}=m/(\gamma\tau)$ quantifying, respectively, the activity and the inertia of the system.


Particles move in a box of size $L$ with periodic boundary conditions and interact through instantaneous collisions that change the velocities of particles $i$ and $j$ from ${\bf v}_i,{\bf v}_j$ to ${\bf v}'_i,{\bf v}'_j$, conserving kinetic energy and total momentum but leaving unchanged $\mathbf{e}_i$ and $\mathbf{e}_j$ (see Fig.~\ref{fig:sketch}(b) and End Matter, EM, for the exact collision rules).
Since collisions conserve kinetic energy, the evolution of $T=m \langle {\bf v}^2 \rangle/d$ in the homogeneous state (no macroscopic drifts and no spatial gradients) is dictated by the following equation obtained by averaging the Ito-transformation of Eq.~\eqref{eq:dynamics_v}:
\begin{equation}
    \label{eq:kinen}
    \dot T = -(2\gamma/m) (T-T_0)+(2\gamma/d)v_0 \langle \mathbf{v} \cdot \mathbf{e}\rangle \,.
\end{equation}
The last term is the so-called {\em active power}, that quantifies the heating efficiency of self-propulsion: the randomizing effect of collisions reduces this term and produces an indirect local cooling of the system, see Fig.~\ref{fig:sketch}(d).

\begin{figure}
    \centering\includegraphics[width=0.99\linewidth]{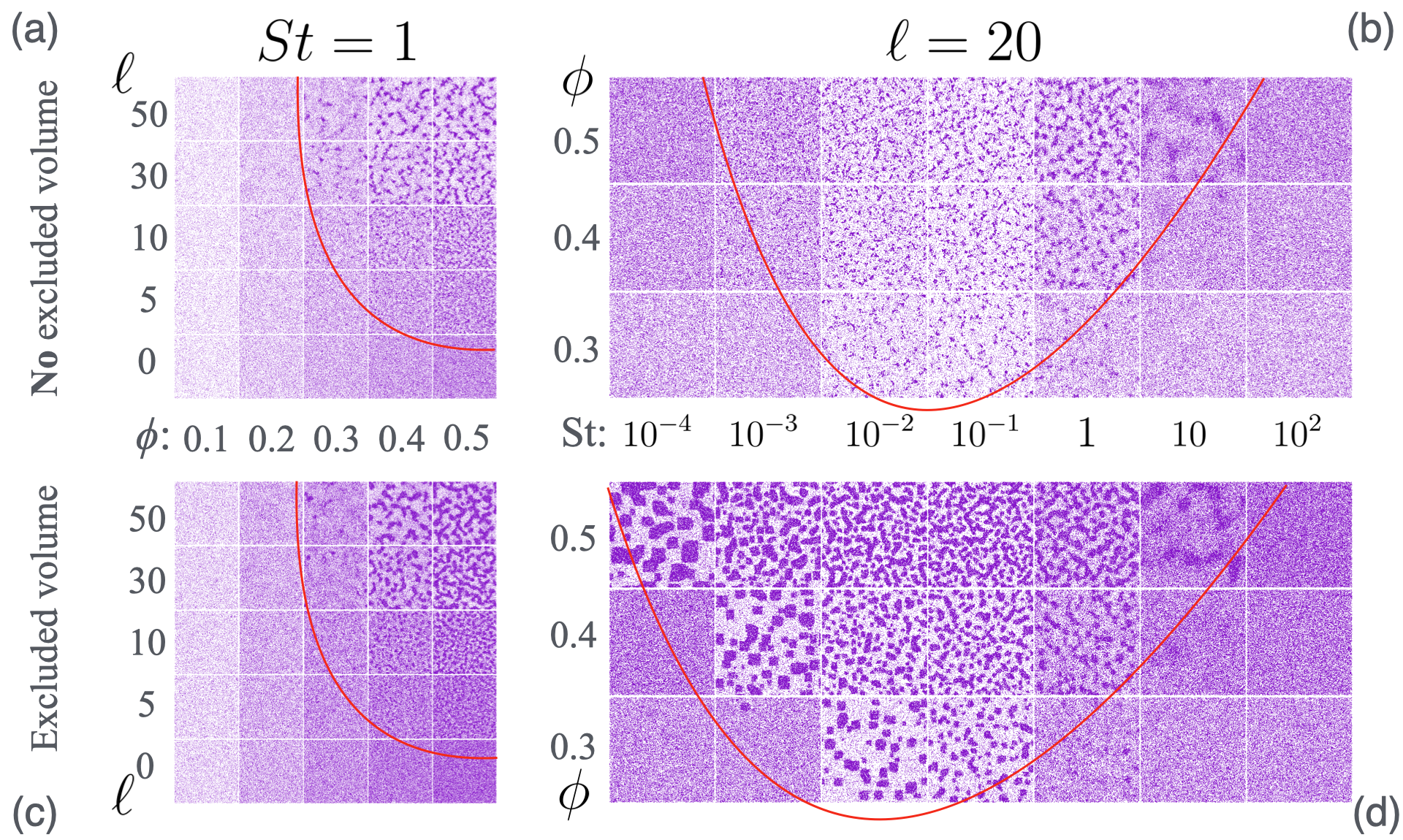}
    \caption{
    Numerical results obtained using ADSMC with the Enskog collision rate.
    (a), (c) Snapshot phase diagrams in the plane of P\'eclet number $\ell$ and packing fraction $\phi$ at Stokes number $St = 1$.
    (b), (d) Snapshot phase diagrams in the $(\phi, St)$ plane for $\ell = 20$.
    Panels (a)–(b) correspond to simulations without excluded-volume interactions, while (c)–(d) include excluded volume. Red lines serve as guides to the eye. The remaining simulation parameter is $T_0 = 1$.
    }
    \label{fig:simulations}
\end{figure}

{\em Active Direct Simulation Monte Carlo (ADSMC) --} 
To explore the effects of the aforementioned cooling mechanism, the model is simulated through the ADSMC, i.e.\ a particle-based Monte Carlo method that accounts for self-propulsion (Fig.\ref{fig:sketch}(c)). Both time and space are discretized by choosing a time-step $dt$ and a cell size $\Delta$ smaller than the minimal time and length scales involved, yet large enough to allow statistical averaging; results are independent of $dt$ and $\Delta$ within a reasonable range. At each time-step, we evolve the free-streaming active trajectory (Eqs.~\eqref{eq:dynamics_v}-\eqref{eq:dynamics_f}  for $d=2$) and compute interactions by performing elastic collisions at random positions within each cell.
The local collision frequency is estimated as proportional to $\chi(\phi)\sigma\phi\overline{v}$, where $\phi=n\pi\sigma^2/4$ is the local packing fraction (with $n$ the local density) and $\overline{v}$ the typical velocity fluctuation inside the cell. The term $\chi(\phi)$ represents a nonlinear correction accounting for positional correlations in nondilute conditions. We consider two choices: (i) $\chi(\phi)=1$, corresponding to pure molecular chaos, and (ii) $\chi(\phi)$ given by the Carnahan–Starling formula for the density correlation at contact. The latter choice is equivalent to an Enskog correction in kinetic theory, which does not explicitly include volume exclusion since collisions occur at random positions within the cell (see EM for details).

{\paragraph*{Inertial MIPS} --}
 Simulations start with homogeneously distributed particle positions and Gaussian-distributed velocities. In some cases, the homogeneous distribution becomes unstable and develops spatial inhomogeneities with clusters, reminiscent of MIPS. This instability is never observed when pure Molecular Chaos is implemented, $\chi(\phi)=1$. In contrast, when the Enskog correction is included (Fig.~\ref{fig:simulations}~(a)), the instability reproduces the standard density–activity phase diagram of MIPS: clusters emerge either by increasing P\'eclet $\ell$, for a fixed value of Stokes $\textrm{St}$, or by increasing the density at large $\ell$. No reentrance is observed when $\ell$ is further increased beyond the instability threshold, coherently with recent studies that attribute it to interaction softness~\cite{mandal2019motility,feng2025theory}. 

 The phase diagram in the plane of inertia and density (at fixed P\'eclet) shows the suppression of MIPS at large inertia (Fig.~\ref{fig:simulations}~(b)), as previously seen in molecular dynamics~\cite{mandal2019motility} and experiments~\cite{caprini2024dynamical}. Interestingly, MIPS also appears to vanish in the overdamped limit. This is however an effect due to the absence of volume-exclusion: when excluded volume (EV) is switched on in the algorithm (see EM for details), MIPS at small inertia is restored (Fig.~\ref{fig:simulations}~(c)-(d)). Moreover, simulations with EV display multiple clusters of finite physical size, coherently with the binodal decomposition scenario~\cite{su2021inertia}.

Our comparative numerical analysis confirms that volume exclusion is essential in the overdamped regime to recover MIPS, as in previous studies. By contrast, this ingredient is not critical at large inertia, where the MIPS instability is numerically observed without it and, therefore, may arise from a different dynamical mechanism.

\begin{figure}
        {\centering  \includegraphics[width=0.95\linewidth,clip=true]{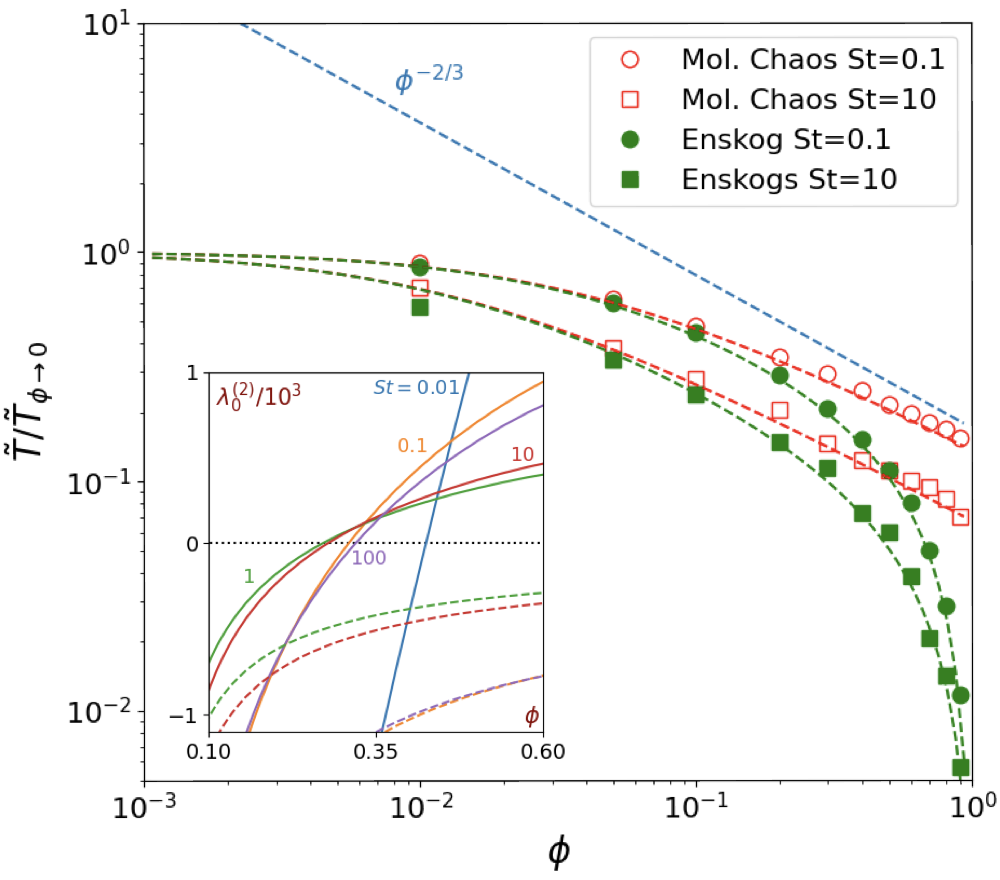}}
    \caption{
    Steady-state temperature $\tilde{T}$, normalized by its dilute-limit value $\tilde{T}_{\phi \to 0}$, as a function of the packing fraction $\phi$ for homogeneous configurations, demonstrating the cooling mechanism. Simulations are performed using both the Molecular Chaos (red) and Enskog (green) collision rates. Colored dashed lines indicate the corresponding theoretical predictions, while the light-blue dashed line serves as a guide to the eye, for the scaling $\tilde{T} \approx z \sim \phi^{-2/3}$.
    Inset: theoretical prediction for the $k^2$-order contribution to the density eigenvalue $\lambda_0^{(2)}$ as a function of $\phi$, for Enskog (solid lines) and Molecular Chaos (dashed lines) collision rates. The colored numbers in the plot denote the Stokes number values.
    In all data, $\ell = 100$ and $\tilde{T}_0 = 2 \times 10^{-4}$. Note that $\mathrm{St}$ is varied by keeping $m = 1$ and $\tau = 1$ fixed while changing $\gamma$. 
    }
    \label{fig:homogeneous}
\end{figure}

{\paragraph{Kinetic theory}--} 
To investigate the dynamical origin of inertial MIPS, we develop a kinetic theory~\cite{baskaran2008transport,brilliantov2010kinetic}. By coarse-graining the  equations of motion, Eqs.~\eqref{eq:dynamics_v}-\eqref{eq:dynamics_f}, we obtain a kinetic equation for the evolution of the single-particle probability density $f=f({\bf r},{\bf v},{\bf e},t)$:
\begin{equation} 
\label{kineq}
\left(    \partial_t+{\bf v} \cdot \partial_{{\bf r}}+(\gamma/m) v_0 {\bf e} \cdot \partial_{\bf v} \right) f=(\mathcal{L}_b+\mathcal{L}_a)f+\mathcal{I}[f_2] \,,
\end{equation}
where the operators $\mathcal{L}_b$ and $\mathcal{L}_a$ represent the interaction with the bath and the active force dynamics, and $\mathcal{I}[f_2]$ describes the inter-particle interactions. The latter term depends on the two-particle probability density $f_2=f_2({\bf r}_1,{\bf v}_1,{\bf e}_1,{\bf r}_2,{\bf v}_2,{\bf e}_2,t)$ (see EM). The crucial assumption in kinetic theories is to close Eq.~\eqref{kineq} by expressing $f_2$ in terms of $f$. Here, consistently with the ADSMC algorithm, we adopt an Enskog–Boltzmann–like ansatz that incorporates density correlations at contact while replacing the hard-sphere collision frequency -- proportional to the relative velocity by a mean collision frequency proportional to the thermal velocity, as in the case of Maxwell molecules. Then, the collision operator takes the nonlinear form $\mathcal{I}[f_2] \to \mathcal{I}_B[f]$ with (see EM)
\begin{align} \nonumber
\mathcal{I}_B[f] = \nu\,
 \int d\mathbf{e}_1  d\mathbf{v}_1 
 d\hat{\boldsymbol{\sigma}}
(
\widehat{b}_{\boldsymbol{\sigma}}
- 
1
)
f(\mathbf{r},\mathbf{v},\e,t)\,
f(\mathbf{r},\mathbf{v}_1,\e_1,t)\,,
\end{align}
where $\nu=\nu(\mathbf{r},t)= (\sigma/2) \sqrt{\frac{T(\mathbf{r},t)}{m}}\chi(\phi(\mathbf{r},t))$ and the operator $\widehat{b}_{\boldsymbol{\sigma}}$ transforms the velocities into the postcollisional velocities (see EM) when acting on functions of $(\vv,\vv_1)$~\footnote{The $1/2$ factor in $\nu$ corresponds to the correct choice for comparison with ADSMC simulations. }. 


{\paragraph{Spatially homogeneous stationary state} --}  In the stationary and spatially homogeneous regime, the kinetic equation for the system admits a solution $f_s({\bf v},{\bf e})$ whose relevant moments can be theoretically evaluated (see EM). Thus, in what follows, averages refer to this particular state.
From here, we can predict the correlation between velocity and orientation, $\langle\mathbf{e}\cdot\mathbf{v}\rangle$ (proportional to the active power), which is $0$ in equilibrium when $v_0=0$.
Out of equilibrium, when $v_0>0$ and $\tau>0$, we get (see EM) a closed equation for $z(n,T_0,v_0,\gamma,\tau,m)=\langle {\bf e} \cdot {\bf v} \rangle/v_0$:
\begin{equation} \label{eqstate2}
    \left[1+\textrm{St}^{-1}+ \sqrt{2}\phi \chi(\phi)  \ell\left(\tilde{T}_0+z\right)^{1/2}\right]z=2\textrm{St}^{-1},
\end{equation}
where we introduce $T_a=m v_0^2/2$ and the rescaled thermal temperature $\tilde{T}_0=T_0/T_a$~\footnote{See EM, Physical Units, for a discussion of the Peclet number.}. In the low density limit, one has $z \to \frac{2}{1+\textrm{St}}$. Note that the knowledge of $z$ gives also the knowledge of the stationary homogeneous temperature $\tilde{T}(\phi,\tilde{T}_0,\ell,\textrm{St})=T/T_a=\tilde{T}_0+ z(\phi,\tilde{T}_0,\ell,\textrm{St})$. Crucially, when $\tilde{T}_0 \ll 1$ (e.g.\ for large $v_0$), one has $\tilde{T} \approx z$ and Eq.~\eqref{eqstate2} depends upon $\gamma$, $m$, $\tau$ and $v_0$ only through $\textrm{St},\ell$: then, in the high density limit $\tilde{T} \to \frac{\sqrt{2}}{( \textrm{St}\, \sigma \, \ell)^{2/3} }  [\phi \chi(\phi)]^{-2/3}$. However, in general the equation depends also on $\tilde{T}_0$ and therefore scalings should take also this parameter into account. We are now in the position of quantifying how the temperature cools down with increasing density, even if the collisions are elastic. Notwithstanding this, when Molecular Chaos is strictly enforced, no phase separation can be observed at odds with the phenomenology of cooling granular gases, where a spatially homogeneous state can be unstable even in the low density limit~\cite{goldhirsch1993clustering}. 
The crucial point is how fast $T$ decays with $\phi$. With cooling granular gases one may have a (rescaled) decay faster than $\phi^{-1}$, while here, with Molecular Chaos $\chi=1$, the decay of the temperature is not faster than $\phi^{-2/3}$. As known since the general analysis by Cates and Tailleur~\cite{cates2015motility}, confirmed below by our theory, this is not sufficient to  observe MIPS. In Fig.~\ref{fig:homogeneous}, we show the decay of $\tilde{T} \approx z$ with Molecular Chaos and with the Enskog correction, comparing it with single-cell (homogeneous) ADSMC simulations, observing an excellent agreement without any adjustable parameter.


\begin{figure}
    \includegraphics[width=0.98\linewidth,clip=true]{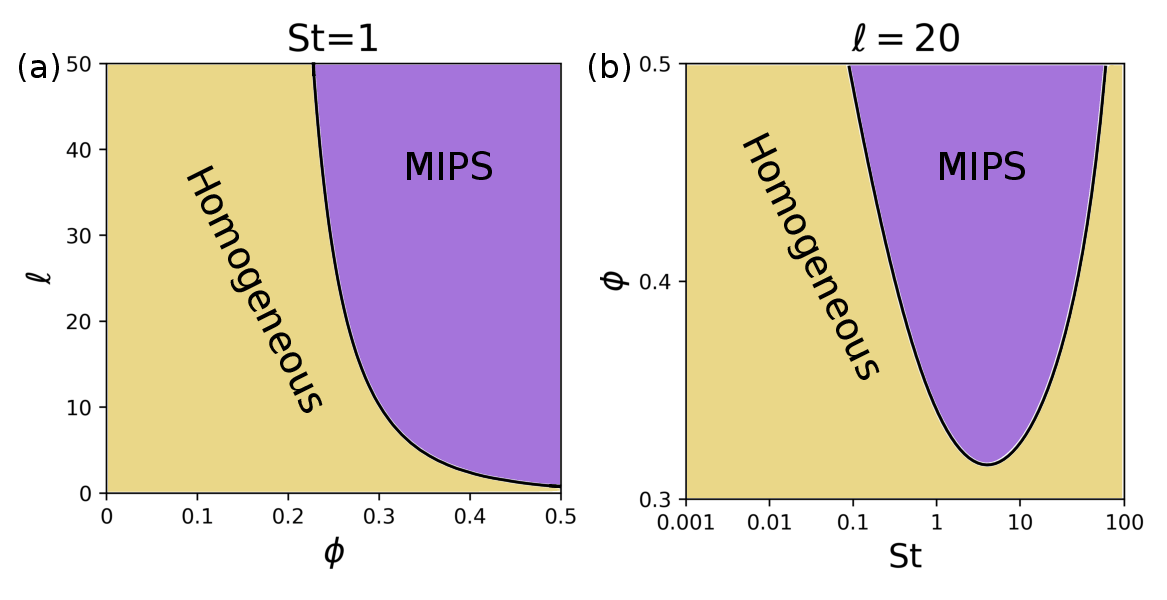}    
\caption{Theoretical phase diagrams from Eq.~\eqref{lambda}. (a) phase diagram in the plane of P\'eclet number $\ell$ and packing fraction $\phi$ at Stokes number $St=1$. (b) Phase diagram in the plane of $\phi$ and $St$ at $\ell=20$. The violet region (MIPS) highlights the region where the homogeneous density is unstable ($\lambda_0^{(2)}>0$).
}
    \label{fig:thdiagram}
\end{figure}

{\em Density linear stability from the kinetic equation --} Deviations of the probability density from the stationary spatially homogeneous solutions are indicated by $\delta f({\bf r},{\bf v},{\bf e},t)=f({\bf r},{\bf v},{\bf e},t)-f_s({\bf v},{\bf e})$. In the Supplemental Material (SM), we show that $\delta f$ obeys
\begin{equation}
    \partial_t \delta f({\bf r},{\bf v},{\bf e},t)=[\Lambda({\bf v},{\bf e})-{\bf v} \cdot \partial_{\bf r}]\delta f({\bf r},{\bf v},{\bf e},t)
\end{equation}
where $\Lambda({\bf v},{\bf e})$ is the homogeneous linearized Boltzmann operator (see SM for its definition). While an eigenvalue problem is defined for each Fourier modes $\delta f_k = \int d{\bf r}e^{-i {\bf k}\cdot {\bf r}}\delta f$, remarkably, we can identify the exact eigenfunction of $\Lambda$ associated to the null eigenvalue. 
By assuming that the eigenvalues and eigenfunctions are analytic in $k$, and focusing on the one that vanishes for $k \to 0$ (that governs the dynamics in the long-time limit), we expand as $\lambda_{0,k}=i k \lambda_0^{(1)}+k^2\lambda_0^{(2)}+O(k^3)$,
where we have set ${\bf k}=k {\hat x}$ by isotropy. As the problem is non-degenerate,  perturbation theory allows us to find $\lambda_0^{(1)}=0$ and
\begin{equation}
    \label{lambda}
        \frac{\lambda_0^{(2)} }{D_0}= -\left(1+\textrm{St}^{-1}\right)\frac{d}{d \phi} [\phi \tilde{T}(\phi,\tilde{T}_0,\ell,\textrm{St})]+\textrm{St}^{-1} \tilde{T}_0
\end{equation}
with $D_0=T_a/\gamma$ (see SM). The eigenvalue $\lambda_0^{(2)}$ needs to be negative to guarantee the linear stability of the homogeneous state. In the hydrodynamic limit $k \to 0, t\to \infty$ with $k^2 t$ finite, the above result implies that the density modes $n_k(t)$ (Fourier modes of the density field) evolve according to the diffusion equation
\begin{equation} \label{diffusion}
    \partial_t n_k(t)=-D_{eff} k^2 n_k(t)
\end{equation}
    with effective diffusion coefficient $D_{eff}=-\lambda_0^{(2)}$. Note that the same result for the stability of the density field (and its ultimate diffusion evolution) can be obtained by expanding the kinetic equations for the hydrodynamic fields through the Chapman-Enskog method. The expressions~\eqref{lambda} and~\eqref{diffusion} are valid also for hard sphere/disks collisional kernel, but explicit expressions for $T$ and $z$ are not analytically available without some kind of approximation of the kinetic equation. As mentioned before, when Molecular Chaos is enforced ($\chi=1$), the decay of $T$ with $n$ is not fast enough to guarantee that $D_{eff}$ becomes negative, i.e.\ instability is never observed. On the contrary, when Enskog correction is introduced ($\chi<1$), MIPS appears for the parameter region such that $D_{eff}<0$ (inset of  Fig.~\ref{fig:homogeneous}). 
    The theory predicts phase diagrams in the plane of P\'eclet number and packing fraction $\phi$ (Fig.~\ref{fig:thdiagram}~(a)) and in the plane of $\phi$ and Stokes number $St$ (Fig.~\ref{fig:thdiagram}~(b)) in fair agreement with those observed in ADSMC simulations (Figs.~\ref{fig:simulations}(a)-(b))
    In the high inertia limit $\textrm{St} \to \infty$, MIPS disappears since $z \to 0$ and $\lambda_0^{(2)} \to -D_0\tilde{T}_0$, in agreement with previous numerical and experimental results~\cite{caprini2024dynamical,mandal2019motility,de2022motility}. 
    As in ADSMC simulations, MIPS is suppressed in the overdamped limit $\textrm{St} \to 0$, because the homogeneous density is stable, $z \to 0$, and $\lambda_0^{(2)} \to -D_0\tilde{T}_0$.
   This artifact is due to the lack of EV in the theory, e.g.\ the collisional contribution to the pressure is neglected. In fact, when  ADSMC is performed with EV, the instability for $\textrm{St} \to 0 $ reappears, as expected for overdamped active systems.
    

{\em Conclusions --} We have shown that inertial MIPS arises from a kinetic temperature decrease with density, in close analogy to the clustering observed in granular materials governed by dissipative interactions~\cite{goldhirsch1993clustering}. However, unlike that case, here the instability requires an Enskog-like correction to the collision rate, going beyond the molecular gas hypothesis. 
This correlation-induced enhancement of collisions couples density, polarization and temperature fields, giving rise to the cooling mechanism responsible for phase separation. These findings are supported by simulations performed through an active direct Monte Carlo method inspired to the Bird algorithm for molecular and granular gases. Our algorithm is faster than molecular dynamics simulations and avoids the critical time-step limitation of previous Monte Carlo approaches~\cite{levis2014clustering,klamser2021kinetic}. As such, it may represent a new paradigm for large-scale active matter simulations.

\begin{acknowledgments}
MM, MIGS and PM acknowledge the support of Grant No. ProyExcel-00505, funded by the Junta de Andalucía, and Grant No. PID2021-126348NB-I00, funded by MCIN/AEI/10.13039/501100011033 and ERDF “A way of making Europe. LC and AP acknowledge funding from the Italian Ministero dell’Università e della Ricerca under the programme PRIN 2022 ("re-ranking of the final lists"), number 2022KWTEB7, cup B53C24006470006.
\end{acknowledgments}

\bibliography{active}

\newpage
\newpage
\newpage
\section*{End Matter}

\subsection{Active Direct Simulation Monte Carlo (ADSMC)}

In the free streaming step the equation of motions are integrated from  $t_i=i \,dt$ to $t_{i+1}=t_i+dt$, ignoring possible interactions, i.e. setting ${\bf f_{i,int}}=0$. For the system under consideration, the free streaming step consists in evolving the three vectors, for each particle, $\{ {\bf r},{\bf v},{\bf e}\}$. In the following, we show results obtained by employing the Euler-Maruyama integration method, including the stochastic terms at order $\sqrt{dt}$.  During the interaction phase, in each cell a "particle-particle" collision step is operated. When excluded volume (EV) is implemented, a second "excluded volume collision" step is also applied.

The particle-particle collisions are treated by first computing the number of expected colliding pairs, according to the following formula~\footnote{Here $[x]$ stands for the average integer part of its real argument $x$, that is $\lfloor x \rfloor + s$ where $s$ is a random variable that can take value $1$ with probability $x-\lfloor x \rfloor$ and $0$ otherwise ($\lfloor x\rfloor$ is the truncated - or floor integer part of $x$).}:
\begin{equation} \label{colnum}
n_{coll} = \left [{\frac{dt}{t_{coll}}}\right ]  \tilde{N}\,,
\end{equation}
where $\tilde{N}$ is the number of particles in the cell.
 The term $t_{coll}$ is the estimated mean free time, determined according to the Enskog-Maxwell-Boltzmann formula $t_{coll} = (\chi(\phi)\sigma n \,\overline{v} )^{-1}$
where $\sigma$ denotes the particle diameter, $n=\tilde{N}/\Delta^2$ represents the local density of particles,  $\phi=n \pi \sigma^2/4$ is the local packing fraction, and $\overline{v}$ corresponds to the typical particle speed in the cell defined as
$\overline{v}=\sqrt{\langle |{\bf v}-\langle {\bf v}\rangle|^2\rangle}$. Here, averages are restricted to the cell (and if $\tilde{N}\le 1$ no collisions are computed). Once the expected number of collisions, $n_{coll}$, has been determined, a cycle is performed for $n_{coll}$ steps: in each step two particles $i,j$ are randomly chosen in the cell and their velocities are updated according to the following elastic collision rule that conserves momentum and kinetic energy:
\begin{align} \label{crule}
{\bf v}_i' &= {\bf v}_i  - (\Delta {\bf v}\cdot \hat{n}) \hat{n}\\
{\bf v}_j' &= {\bf v}_j  + (\Delta {\bf v}\cdot \hat{n}) \hat{n}
\end{align}
where $\Delta{\bf v}={\bf v}_i-{\bf v}_j$ and $\hat{n}=(\cos(\beta),\sin(\beta))$ with $\beta$ extracted from a uniform probability distribution between $0$ and $2\pi$ and prime indicates post-collisional velocities). 
The crucial assumption of this integration scheme is that particle positions within a cell are ignored when evaluating interactions. This is equivalent to performing a Monte Carlo approximation of the collisional integral $I[f|f]$. By setting $\chi = 1$, one recovers the ADSMC scheme under Molecular Chaos, which, for passive particles, is known to converge to the solution of the Boltzmann equation. In our implementation, we employ the Carnahan–Starling approximation, $\chi = (1 - 7 \phi/16)/(1 - \phi)^2$.


The excluded volume affects every particle of the cell that - during its previous movement - has moved to a different cell where there is already a number of particles $\tilde{N}_{max}$ (e.g.\ in two dimensions we use $\tilde{N}_{max}=\lceil \Delta^2/(\pi \sigma^2/4) \rceil$): in that case the particle is ''bounced back'' to a uniformly random position of the cell of origin,  with a uniformly random rotation of the velocity vector. 

\subsection{Virtual number of particles}

The choice of small $\Delta$ (usually of the order of a few diameters) leads to important finite size effects in the interaction step, in particular too frequent repeated collisions, leading to discrepancies with the kinetic theory. Another finite size effect is the possibility of unrealistic local fluctuations of the packing fraction, with $\phi>1$: even if this happens rarely, it affects drastically the implementation of the Carnahan-Starling $\chi(\phi)$ recipe which diverges at $\phi=1$. 

In the usual application of the DSMC~\cite{bird1994molecular,cercignani2013mathematical}, the number of simulated particles does not correspond to the real physical number. For molecular system this "virtual number" is much smaller than the (typically huge) physical number. In granular gas applications, on the contrary finite size effects are usually dealed with by considering a  larger number of virtual particles, $N_v=\beta N$ of particles, with $\beta>1$. This is our choice for ADSMC. The interaction step is then performed by considering the physical (not virtual) local packing fraction $\phi/\beta$ for the purpose of computing the local collision time $t_{coll}$,: this means that each particle collides a number of times according to the physical and not the virtual local density. The large number of virtual particles however reduces spurious repeated collisions and nonphysical density fluctuations.

In the results of Fig.~\ref{fig:simulations} we have used $\beta=5$. When $\beta=1$ is used the ADSMC cannot be implemented without EV because of the aforementioned problem of packing fraction fluctuations. Even when $\beta>1$ the simulations without EV can present (very rare) fluctuations with local $\phi>1$, we have coped with this problem by regularizing $\chi(\phi)$, i.e. by using the Carnahan-Starling formula for $\phi \le \phi^*$ and $\chi(\phi)=\chi(\phi^*)$ with $\phi^*=0.95$.

\subsection{Physical units}

There are six parameters: $m,\tau,\gamma,T_0,v_0,\sigma$. We have defined three adimensional parameters: $\tilde{T}_0=T_0/(m v_0^2/2)$, $\textrm{St}=m/(\tau \gamma)$, $\ell=\tau v_0/\sigma$. If the three fundamental units for time, length and energy are taken to be $\tau=1$, $\sigma=1$ and $T_0=1$, then the knowledge of the three adimensional parameters gives immediate knowledge of the three dimensional non-unitary parameters: $v_0=\ell$, $m=2/(\tilde{T}_0 \ell^2)$, $\gamma=2/(\tilde{T}_0 \ell^2 \textrm{St})$. With this way of defining units, changes in $\textrm{St}$ are equivalent to changes in $\gamma$.
Note that $\ell=\tau v_0/\sigma$ can be considered a Peclet number: it is in fact (apart from a factor $2$) equivalent to the ratio between the rescaled ballistic time $(\sigma/v_0)/\tau$ and the rescaled diffusion time $(\sigma^2/D_a)/\tau$ with $D_a=T_a \tau$.

\subsection{Explicit expressions for the kinetic operators and equations}
\label{subsec:exp_exp}
The explicit expressions for the kinetic operators of Eq.~\eqref{kineq}, in the main text, are
\begin{align}
\mathcal{L}_b f(\mathbf{r},\mathbf{v} ,\e,t) &=  \frac{\gamma}{m}\partial_\vv \cdot \left(\frac{T_0}{m} \partial_\vv + \bm{v} \right) f(\mathbf{r},\mathbf{v},\e,t)  , \label{op:L_b}\\
\mathcal{L}_a f(\mathbf{r},\mathbf{v} ,\e,t)   &= \frac{1}{\tau} \partial_\e \cdot \left( \e + \partial_\e \right) f(\mathbf{r},\mathbf{v} ,\e,t)  
 \label{op:L_a}, 
 \end{align}
 \begin{multline}
\mathcal{I}[f_2] = \sigma \int d \e_1 \int d \vv_1 \int d \un  |\mathbf{g}\cdot \un |  \times \\ \times \left[ \Theta(\mathbf{g} \cdot \un )\bop -   \Theta(-\mathbf{g} \cdot \un )\right]  f_2(\mathbf{r},\vv,\e,\mathbf{r}+\boldsymbol{\sigma},\vv_1,\e_1,t)
\label{eq:app_If2}
\end{multline}
where ${\mathbf{g}} \equiv {\vv}_{1} - {\vv}$ is the relative velocity, $\bm{\sigma}$ is a vector joining the two particles at contact (pointing away from the particle with velocity $\vv$), and $  \un \equiv \bm{\sigma}/\sigma  $. The operator $\bop$ transforms precollisional velocities $(\mathbf{v},\mathbf{v}_1)$ into their postcollisional values, i.e., $\bop g(\vv,\vv_1) = g (\bop \vv,\bop \vv_1)$, with $\bop \vv = \vv + (\vb{g}\cdot \un)\un $,  $\bop \vv_1 = \vv_1 - (\vb{g}\cdot \un)\un $.

Assuming molecular chaos for precollisional velocities while retaining positional correlations, the two-particle distribution function can be approximated as
\begin{equation}
\label{eq:app_f2}
f_2(\mathbf{r},\vv,\e,\mathbf{r}+\boldsymbol{\sigma},\vv_1,\e_1,t)  =  \chi [\phi(\vb{r},t)] f(\vb{r},\vv,\e, t ) f(\vb{r}+\boldsymbol{\sigma},\vv_1,\e_1,t) ,
\end{equation}
where $\chi[\phi]$ is the density dependent pair correlation function at contact. If Eq.~\eqref{eq:app_f2} is substituted in Eq.~\eqref{eq:app_If2}, we obtain the Enskog collision operator
\begin{multline} \label{eq:enskog_colop}
\mathcal{I}_E[f] = \sigma \int d \e_1  \int d \vv_1 \int d \un |\vb{g}\cdot \un | \chi[\phi] \left[ \Theta(\mathbf{g} \cdot \un )\bop -  \Theta(-\mathbf{g} \cdot \un )\right] \\ \times f(\vb{r},\vv, \e ,t ) f(\vb{r}+\boldsymbol{\sigma},\vv_1,\e_1,t) . 
\end{multline}
Finally, if the single-particle distribution varies weakly over distances of order $\sigma$, and assuming a velocity-independent collision frequency proportional to the local thermal speed--as in the simulations--one may replace $|\vb{g}\cdot\un| \to (1/2) \sqrt{\frac{T(\vb{r},t)}{m}}$ in Eq.~\eqref{eq:enskog_colop}.
Under this approximation, the form $\mathcal{I}_B[f]$ in the main text is recovered.

\subsection{Homogeneous stationary state}

To analyze the stationary state, we define the global moments in this state.
For a microscopic function $\psi(\mathbf{v},\e)$, its average is defined as
\begin{align}
\langle \psi(\mathbf{v},\e) \rangle
\equiv  \frac{1}{n} \int d\e \int d\mathbf{v} \psi(\mathbf{v},\e) f_s(\mathbf{v},\e),
\end{align}
 where $f_s(\mathbf{v},\e)$ is the homogeneous stationary distribution function of the Boltzmann-Fokker-Planck equation. 

To characterize the stationary state, we derive equations for the mixed and orientational correlators, $\langle e_i v_j \rangle$, $\langle e_i e_j \rangle$ and stationary temperature $T \equiv \frac{m}{2} \langle v^2 \rangle$. This is done by multiplying Eq.~\eqref{kineq}, evaluated at the stationary distribution $f_s$ (so that $\partial_t f_s = \partial_{\mathbf{r}} f_s = 0$), by $e_i v_j$, $e_i e_j$ and $v^2$, respectively, and integrating over $\mathbf{e}$ and $\mathbf{v}$. In this procedure, the collision integral is treated under the Enskog–Boltzmann molecular chaos assumption. Together with the stationary temperature condition, this procedure yields
\begin{subequations}
\begin{align}
&\langle e_i e_j \rangle = \delta_{ij},\\
&T=T_0 + \frac{m}{2}v_0 \langle \e \cdot \vv   \rangle,\\
& \langle e_i v_j \rangle \left[ \frac{\gamma}{m} + \tau^{-1} +  \frac{\pi}{2} n \sigma \chi(n)     \sqrt{T/m} \right] = \frac{2\gamma }{m} v_0 \delta_{ij} . \label{eq:ev1}
\end{align} 
\end{subequations}
Using the relation between packing fraction and density, Eq.~\eqref{eq:ev1} simplifies to $\langle e_i v_j \rangle = 0$ for $i \neq j$, and
\begin{align}
\left[ \frac{\gamma}{m} + \tau^{-1} + \frac{\sqrt{2} \phi \chi(\phi)}{\sigma}     \left(\frac{2 T_0}{m}+v_0\langle \e \cdot \vv \rangle \right)^{1/2} \right] \langle \e \cdot \vv  \rangle = \frac{2\gamma }{m} v_0 ,
\end{align}
from which Eq.~\eqref{eqstate2} directly follows. 

By direct derivation of Eq.~\eqref{eqstate2} in the homogeneous state, we get an equation for $y= \frac{d}{d \phi} ( \phi \tilde{T} )$ which is central for the knowledge of the density eigenvalue $\lambda_0^{(2)}$, see Eq.~\eqref{lambda}. This equation is linear in $y$, leading immediately to 
\begin{align}
\frac{d}{d \phi } ( \phi \tilde{T} ) = - \frac{\sqrt{2}\ell \,\text{St} \, z \left[ \phi \chi'(\phi) + \chi (\phi)  \right] (z+\tilde{T}_0)}{(\sqrt{2}/2)\ell \phi \,\text{St}\,  \chi(\phi) (3z+2 \tilde{T}_0) + (\text{St}+1)\sqrt{z+\tilde{T}_0}} . 
\end{align}

\end{document}